\documentclass[twocolumn,superscriptaddress,showpacs,pra]{revtex4}
\usepackage{epsfig}
\usepackage{delarray}
\usepackage{amsmath}
\usepackage{bm}
\usepackage{epstopdf} 

\newcommand{\ket}[1]{| \, #1 \rangle}
\newcommand{\bra}[1]{ \langle #1 \,  |}

\begin{document}

\title{Photonic multipartite entanglement conversion using nonlocal operations}

\author{T. Tashima}
\affiliation{Department f\"ur Physik, Ludwig-Maximilians-Universit{\"a}t, D-80799 M{\"u}nchen, Germany}
\affiliation{Max-Planck-Institut f{\"u}r Quantenoptik, Hans-Kopfermann-Strasse 1, D-85748 Garching, Germany}
\affiliation{Division of Materials Physics, Department of Materials Engineering Science, Graduate School of Engineering Science, Osaka University, Toyonaka, Osaka 560-8531, Japan}
\affiliation{School of Chemistry and Physics, University of KwaZulu-Natal, Durban 4001, South Africa}
\email{ttoshi@oregano.ocn.ne.jp}

\author{M. S. Tame}
\affiliation{School of Chemistry and Physics, University of KwaZulu-Natal, Durban 4001, South Africa}
\affiliation{National Institute for Theoretical Physics, KwaZulu-Natal, South Africa}
\email{markstame@gmail.com}

\author{\c{ S}. K. \"Ozdemir}
\affiliation{Department of Electrical and Systems Engineering,
Washington University in St. Louis, St. Louis, MO 63130 USA}
\email{ozdemir@wustl.edu}

\author{F. Nori}
\affiliation{Center for Emergent Matter Science, RIKEN, Saitama 351-0198, Japan}
\affiliation{Physics Department, University of Michigan, Ann Arbor, Michigan 48109-1040,
USA}
\author{M. Koashi}
\affiliation{Photon Science Center, The University of Tokyo, Bunkyo-ku, 113-8656, Japan}

\author{H. Weinfurter}
\affiliation{Department f\"ur Physik, Ludwig-Maximilians-Universit{\"a}t, D-80799 M{\"u}nchen, Germany}
\affiliation{Max-Planck-Institut f{\"u}r Quantenoptik, Hans-Kopfermann-Strasse 1, D-85748 Garching, Germany}

\date{\today}
\begin{abstract}
We propose a simple setup for the conversion of multipartite entangled
 states in a quantum network with restricted access. The scheme uses
 nonlocal operations to enable the preparation of states that are
 inequivalent under local operations and classical communication, but
 most importantly does not require full access to the states. It is based on a flexible linear optical conversion gate that uses photons, which are ideally suited for distributed quantum computation and quantum communication in extended networks. In order to show the basic working principles of the gate, we focus on converting a four-qubit entangled cluster state to other locally inequivalent four-qubit states, such as the GHZ and symmetric Dicke state. We also show how the gate can be incorporated into extended graph state networks, and can be used to generate variable entanglement and quantum correlations without entanglement but nonvanishing quantum discord.
\end{abstract}
\maketitle


\section{Introduction}
Entanglement between two or more particles is an elementary
resource for a variety of quantum communication and computing
tasks~\cite{Hor,Gis,s4,clusterrev, mutli}. Recently, sophisticated
quantum network architectures relying on resources with different entanglement
structures among the nodes of a network have been proposed for
distributed quantum communication and
computing~\cite{Kimble,Chir,Per}. In order to make full use of the
nodes of a network, it is important to identify and to prepare the
optimal shared resource for a given task, as well as to understand the
equivalence relations among different entangled resource states and the
inequivalent classes of entanglement. 

Progress in the general study of multipartite entanglement was stimulated
by the finding that entangled states of three qubits cannot be converted into each other by local operations and classical communication (LOCC)~\cite{s6,s6b}. While for bipartite pure states there is only one class of states under LOCC, the situation changes dramatically for more qubits. 
For example, for four parties,
Greenberger-Horne-Zeilinger (GHZ) states~\cite{s5}, W states~\cite{s6,
s7, s8}, Cluster states~\cite{s4,clusterrev} and Dicke states~\cite{s10}
come from inequivalent entanglement classes~\cite{s25,cschmid}.
\begin{figure}[t]
\includegraphics[scale=1]{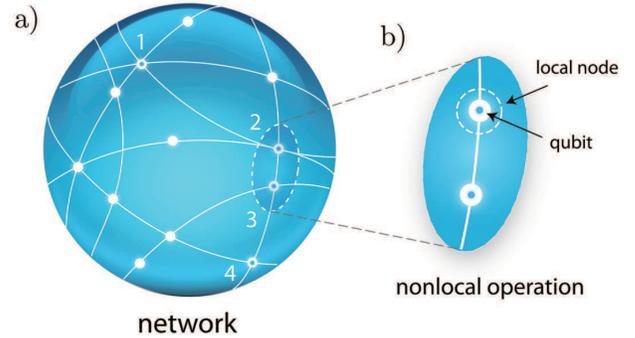}
\caption{Quantum network with restricted access. We consider a nonlocal conversion gate that operates on qubits from two local nodes of the network, as shown in panel (b), and enables the conversion of different types of multipartite entangled states for quantum networking applications. While we focus on four-qubit entangled states in our study, the overall principle could be extended to larger more complex networks due to the operating range of the conversion gate. 
\label{fig:0s}}
\end{figure}

This classification led to a series of efforts to find methods and schemes to prepare and
characterize these states in different physical setups~\cite{s6e}, ranging from
photonics~\cite{mutli} and nuclear magnetic resonance~\cite{NMR}, to ion
traps~\cite{Ion, Super} and superconducting circuits~\cite{Super0, Super1}. Studies have also
focused on forming complete toolboxes of physically realizable
operations and gates to manipulate, expand and fuse entangled states of
a certain type to form larger states of the same type such that
entangled networks can be formed~\cite{cluster1,
cluster2,s11,s12,s13,s15,s16,s22,
s17,s18,s19,s20,s21,ikuta, s23,s24,WWiec,WWiec1,Prevedel,Sinan,Fatih}. In parallel to this
work, there has also been much interest in the use of entangled states
in quantum information processing tasks, especially in finding tasks for
which the states from one entanglement class may be more efficient than
one from another class. For example, cluster states and graph states are
universal resources for quantum computing~\cite{s4,clusterrev}, GHZ
states have been proposed as resources for achieving consensus in
distributed networks without classical post-processing~\cite{GHZWdist}
and W states have been proposed as resources for leader election in
anonymous quantum networks~\cite{GHZWdist} and asymmetric
telecloning~\cite{clone}. From a
network perspective, the qubits are distributed to the nodes where the users have access only to a limited
number of qubits. Typically they make use of classical communication channels to perform individually or collectively an assigned
task ~\cite{GHZWdist,clone,s1}. In some cases, two nodes of a network may
be close enough to each other such that joint operations can be carried
out with a small overhead in communication, as shown in Fig.~\ref{fig:0s}. Such a scenario could also describe the case where a node
of a network is also the node of another network and holds two or more
qubits belonging to different networks. This `nonlocal' node could
manipulate qubits from different networks for the purpose of fusing
those networks into a larger merged network. This and similar tasks are
of considerable importance as they help us to understand how to
efficiently exploit multipartite entanglement for quantum communication protocols and how to design practical applications.

Recent work has focused on the
transformation of different types of entangled states into each other by
local one and two-qubit operations while allowing classical
communication among all the nodes. For example, Kiesel~{\it
et~al.}~\cite{D4} demonstrated that a four-partite symmetric Dicke state
can be used to prepare a W state among three parties if one of the
parties in the network projects their qubit onto the computational
basis. The initial symmetric Dicke state can be used in certain quantum
versions of classical games whereas the final W state cannot be used \cite{clone1,clone2}. On the other hand, the final W state can be used for asymmetric telecloning of a qubit \cite{clone}. Along the same lines, another interesting work is
that of Walther~{\it et~al.}~\cite{s24.1}, who showed that by the
application of appropriate generalized measurements on a tripartite GHZ
state, one can obtain a state whose fidelity to a tripartite W state
approaches one, when the probability of success approaches
zero. Tashima~{\it et~al.}~\cite{s17}, on the other hand, have proposed
a series of optical gates that can prepare W states of arbitrary size
by accessing only one qubit of the W state. They also introduced a
scheme in which two parties sharing a pair of Bell states can prepare
tripartite W states and expand them to larger sizes with the help of
ancillary qubits, again by accessing only one qubit in the network \cite{s17,s18,s21}.
\begin{figure}[t]
\includegraphics[scale=1]{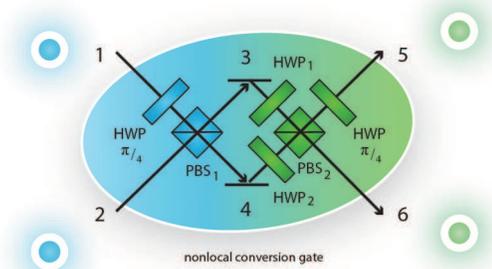}
\caption{Nonlocal gate using linear optics. The gate is used in our
 restricted-access network scenario to convert between different types of
 entanglement structures.
\label{fig:1s}}
\end{figure}

In this work we go beyond those concepts by introducing a scheme that uses `bilocal' operations to enable the preparation of resource states in a network that are inequivalent under
LOCC, but crucially does not require full access to all of the resource state's constituent elements. 
We call the operations bilocal, as they involve a nonlocal two-qubit gate and local one-qubit operations. Here, the two qubits operated on by the nonlocal gate need to be in close proximity, whereas the other qubits do not. This is in contrast to a fully nonlocal operation that would require all qubits to be in close proximity for it to be applied.
Our scheme employs a flexible linear optical conversion gate. The gate can be
used to generate variable entanglement, as well as quantum correlations
without entanglement captured by the quantum discord. While we focus on four-qubit entangled states in our study,  {\it e.g.} cluster, GHZ and
Dicke states, the overall principle can be extended to larger and more complex networks due to the flexible operation of the gate. 
We highlight this extension briefly in our work.

The paper is organized as follows: In Sec.~II, we describe the
principles of the nonlocal gate used in our conversion scheme and
highlight its overall range of operation. We also show how it can
prepare states that are either entangled or have nonzero discord. In Sec. III, we show that the
gate can be used to convert a four-qubit cluster state into other
four-qubit entangled states by operating nonlocally on only two qubits. 
In Sec. IV we show some examples of incorporating the nonlocal gate in an extended graph state network. Finally, in Sec. V we provide a brief summary and outlook.
%


\section{Nonlocal gate for photonic state conversion}
Let us start by introducing the nonlocal two-qubit gate for
photonic state conversion. The gate shown in Fig.~\ref{fig:1s} is composed of two polarizing beamsplitters (PBSs) and four half-wave plates (HWPs). At the most basic level it functions as a tunable polarization-dependent beamsplitter (PDBS). 
However, its advantage over the standard PDBS already used in experiments~\cite{s18,Kiesel05,Clark09} is that as the operation of the individual components can be adjusted easily the configuration is far more flexible and readily implemented experimentally. This is in direct contrast to a bulk PDBS, whose operation is set once fabrication has been completed and cannot be changed. The flexibility of this tunable PDBS complements well the work on a tunable polarization-independent beamsplitter~\cite{Ma11}. 
In this section we consider the individual components of the gate in order to derive the necessary expressions to demonstrate its working principles and range of operation. We then discuss how it can be used to generate entanglement and more general quantum correlations. 


\subsection{Working principle of the conversion gate}
The conversion gate is shown in Fig.\ \ref{fig:1s}. Its successful
operation is based on postselection such that one photon is detected in each of the output modes, labeled 5 and 6. The Kraus operator $E_0$ of the gate transforms the input state
$\rho_{in}=\ket{\psi_{in}}\bra{\psi_{in}}$ according to
\begin{eqnarray}
\rho_{in} \to \frac { E_0\ket{\psi_{in}}\bra{\psi_{in}}E_0^\dagger}{p_{s}} \label{eq:12}
\end{eqnarray}
where $p_{s}={\rm Tr}(E_0\ket{\psi_{in}}\bra{\psi_{in}}E_0^\dagger)$
is the success probability. Here, the action of the HWPs in modes 3 and 4 for the polarization degree of freedom of one photon is given by
\begin{eqnarray}
\ket {H}_j &\rightarrow& \cos(2\theta_l)\ket {H}_j+\sin(2\theta_l)\ket {V}_j \nonumber \\
\ket {V}_j &\rightarrow& \sin(2\theta_l)\ket {H}_j-\cos (2\theta_l)\ket {V}_j. \label{eq:a9}
\end{eqnarray}
where $j=3,4$ labels the modes of the HWPs and $l=1,2$ labels the HWPs. When the input state is $\ket{HV}_j$, the action of the HWPs in
modes 3 and 4 is given by
\begin{eqnarray}
\ket {HV}_j &\rightarrow& \sqrt 2~\cos(2\theta_l)\sin(2\theta_l)\ket {2H}_j\nonumber \\&&-(\cos^2(2\theta_l)-\sin^2(2\theta_l))\ket {HV}_j\nonumber \\&&+\sqrt 2~\sin(2\theta_l)\cos (2\theta_l)\ket {2V}_j,\label{eq:a9_1}
\end{eqnarray}
where we have used the definition $\ket{HV}_j$ for one horizontal polarized photon and one vertical polarized
photon in the same mode $j$, and $\ket
{2H}_j$ ($\ket
{2V}_j$) for two horizontal (vertical) polarized photons in mode $j$. 
In addition, the first PBS applies the unitary transformations:
$\ket{H}_1 \to \ket{H}_4$, $\ket{H}_2 \to \ket{H}_3$, $\ket{V}_1 \to
\ket{V}_3$ and $\ket{V}_2 \to \ket{V}_4$, and the second PBS applies
similar transformations: $\ket{H}_3 \to \ket{H}_6$, $\ket{H}_4 \to
\ket{H}_5$, $\ket{V}_3 \to \ket{V}_5$ and $\ket{V}_4 \to
\ket{V}_6$. By combining all the unitary operations of the HWPs
and PBSs, the Kraus operator of the nonlocal gate can be written as
\begin{eqnarray}
E_0 &=& (\alpha_1-\beta_1)\ket {HH}_{56}~_{12}\bra {HH}\nonumber\\&& +(\alpha_2-\beta_2)\ket {VV}_{56}~_{12}\bra {VV}\nonumber\\&& + \mu_1\ket {HV}_{56}~_{12}\bra {HV} - \mu_2\ket {VH}_{56}~_{12}\bra {HV}\nonumber\\&& + \mu_1\ket { VH}_{56}~_{12}\bra {VH}- \mu_2\ket {HV}_{56}~_{12}\bra {VH}   \label{eq:10}
\end{eqnarray}
where $\alpha_l=\cos^2(2\theta_l)$, $\beta_l=\sin^2(2\theta_l)$,
$\mu_1=\cos(2\theta_1)\cos(2\theta_2)$ and
$\mu_2=\sin(2\theta_1)\sin(2\theta_2)$. The Kraus operator $E_0$ corresponds to a
successful gate operation with probability $p_s$, where each of the
output ports (modes 5 and 6) has one photon. It is clear
that by tuning the rotation angles of $\rm HWP_1$ and  $\rm HWP_2$, the gate acts as a
tunable PDBS, enabling the construction of different Kraus operators. This makes it possible to perform different tasks using this
simple linear optical construction. 
\begin{figure}[t]
\includegraphics[scale=1]{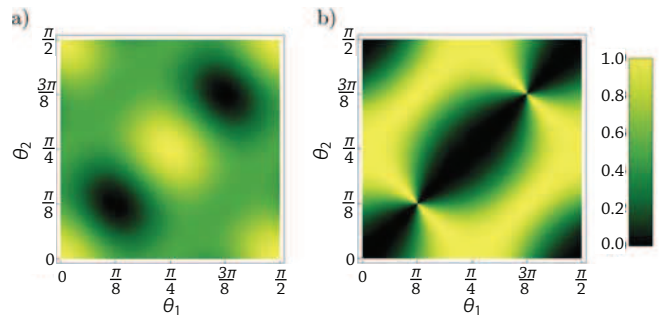}
 \caption{ {\bf (a):}~Success probability for the input state
 $\ket{\psi_{in}}=\ket{+}\ket{+}$ for the conversion gate operating with
 HWPs as polarization rotators at angles $\theta_1$ and $\theta_2$. {\bf
 (b):}~Entanglement generated from this input as quantified by the
 concurrence.
}\label{fig:3s}
\end{figure}


\subsection{Entanglement generation}
We now show some basic examples that demonstrate the entangling power of the
conversion gate and its PDBS ability. In Fig.~\ref{fig:3s}, we show the
success probability and a range of states with different amounts of
entanglement generated by tuning the angles $\theta_1$ and $\theta_2$, as quantified by the
concurrence~\cite{Conc}. Here, the input state is
$\ket{\psi_{in}}=\ket{+}\ket{+}$, where $\ket{+} = (\ket {H}+\ket {V})/\sqrt 2$.
 Applying the Kraus operator $E_0$ of Eq.~(\ref{eq:10}) on this input state yields
\begin{eqnarray}
E_0\ket{\psi_{in}}&=&\frac{1}{2}\big[(\alpha_1-\beta_1)\ket {HH}+(\alpha_2-\beta_2)\ket {VV}\nonumber\\
&& +\sqrt{2}(\mu_1-\mu_2)\ket{\Psi^+}\big] \label{eq:12gg}
\end{eqnarray}
where $\ket{\Psi^+} = (\ket{HV}+\ket{VH})/\sqrt{2}$.

First, in order to prepare the Bell state $\ket{\Psi^+}$ from the state
space of the output in Eq.~(\ref{eq:12gg}) we need to set
$\alpha_1-\beta_1=0$ and $\alpha_2-\beta_2=0$ by rotating the angles of
the HWPs in modes 3 and 4 as $\theta_1=(2k+1)\frac{\pi}{8}$ and
$\theta_2=(2n+1)\frac{\pi}{8}$, respectively, for $k$ and $n$ taking
non-negative integer values. The success probability is maximized when
$\mu_1-\mu_2=\cos2(\theta_1+\theta_2)=\mp 1$, leading to
$\theta_1+\theta_2=m\pi/2$ for some integer $m$. These three equations
are satisfied simultaneously when
$(2k+1)\frac{\pi}{8}+(2n+1)\frac{\pi}{8}=m\frac{\pi}{2}$. This can be reformulated to $k+n+1=2m$, implying that $k+n$ should be an odd number. In
other words, when $k$ is even $n$ is odd and vice versa. Thus the
solution we are looking for is $\mu_1=-\mu_2=\pm 1/2$ resulting in
$E_0=\pm\ket {\Psi^+}\bra {\Psi^+}$. Consequently, the Kraus operator $E_0$ prepares the output state $\rho_{out}=\ket{\Psi^+}\bra{\Psi^+}$ with the success probability $p_s=1/2$. 
In order to give an idea of the robustness of the success probability to variations in the optical components, as in an experimental implementation, we performed a Monte Carlo simulation, varying the angles $\theta_1$ and $\theta_2$ of the wave plates around their ideal values. We set a range of variation for the angles of $\pm 10$\% from the ideal value. Performing 5000 runs (to reach asymptotic behaviour) we find an average success probability of $p_s=0.509 \pm 0.010$, which clearly shows that the success probability is quite robust to realistic perturbations.

Similarly, we see from Eq.~(\ref{eq:12gg}) that if we were able to set
$\mu_1-\mu_2=0$ and $\alpha_1-\beta_1=\pm(\alpha_2-\beta_2)$ this would
prepare the Bell state
$\ket{\Phi^\pm}=[\ket{HH}\pm\ket{VV}]/\sqrt{2}$. For the former
equality, we find $\theta_1+\theta_2=(2m+1)\frac{\pi}{4}$ for
non-negative integer numbers  $m$. The latter equality is satisfied when
$4\theta_2=4\theta_1+2n\pi$ and $4\theta_2=4\theta_1+(2k+1)\pi$,
respectively for $+$ and $-$ signs. It is easy to show under these
conditions that the case with the $+$ sign gives
$\alpha_1-\beta_1=\alpha_2-\beta_2=0$ which is not desirable. On the
other hand, the case with the $-$ sign leads to
$\theta_1=(m-k)\frac{\pi}{4}$ and $\theta_2=(m+k+1)\frac{\pi}{4}$ for
which $\alpha_1-\beta_1=-(\alpha_2-\beta_2)$ is always satisfied. The
above implies that we can prepare only the state $\ket{\Phi^{-}}$ with
the success probability $p_s=1/2$. In this case, we obtain
$\alpha_1=\beta_2=\mu_1=\mu_2=0$ and $\alpha_2=\beta_1=1$, leading to
$E_0= \ket {HH}\bra {HH}-\ket {VV}\bra {VV}$ and the output state
$\rho_{out}=\ket{\Phi^-}\bra{\Phi^-}$ with success probability
$p_s=1/2$. 
A Monte Carlo simulation with a $\pm 10$\% variation on the wave plate angles gives a success probability of $p_s=0.481 \pm 0.021$.

It is clear that starting with the input state $\ket{\psi_{in}}=\ket{+}\ket{+}$ €€(or $\ket{\psi_{in}}=\ket{-}\ket{-}$, where $\ket{-} = (\ket {H}-\ket {V})/\sqrt 2$)), the proposed gate
can prepare a range of states with various amounts of entanglement as seen in Fig.~\ref{fig:3s}, including separable and maximally entangled states, by simply setting the rotation angles of $\rm HWP_1$ and  $\rm HWP_2$ correctly. 
It is also noted that the output state resulting from the input state $\ket{\psi_{in}}=\ket{+}\ket{-}$ or $\ket{\psi_{in}}=\ket{-}\ket{+}$ can be obtained from the input state $\ket{\psi_{in}}=\ket{+}\ket{+}$ by locally compensating the phase shift in either mode 5 or 6. 


\subsection{Discord generation}

\begin{figure}[t]
\includegraphics[scale=0.9]{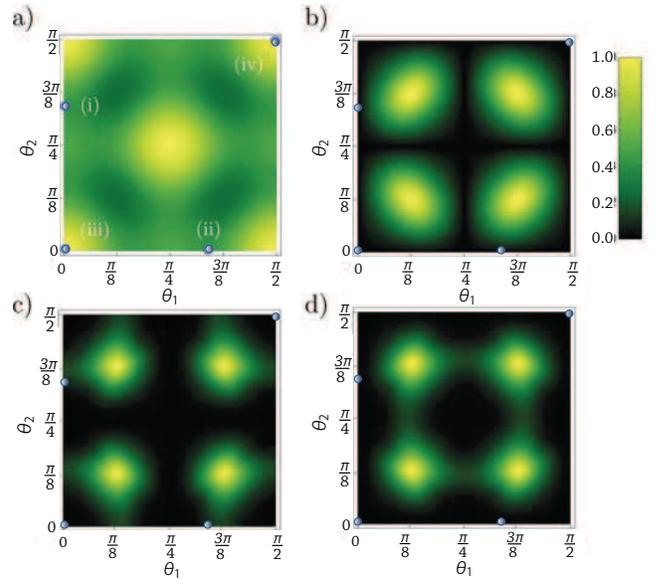}
\caption{Generation of nonclassical correlations by the conversion gate that are not due to entanglement. {\bf (a):} Success probability of the conversion gate. Here points (i) and (ii) correspond to the output state $\frac {5}{7}\ket {H}\bra{H} \otimes \ket{\phi}\bra{\phi}+\frac {2}{7}\ket {V}\bra{V} \otimes \ket{+}\bra{+}$, where $\ket{\phi}=(2\ket{H}-\ket{V})/\sqrt 5$. Points (iii) and (iv) correspond to the output state $\frac {1}{2}\openone \otimes \ket{+}\bra{+}$. {\bf (b):} Entanglement generated as quantified by the concurrence. {\bf (c):} Discord generated with respect to measurements performed on the second qubit. {\bf (d):} Discord generated with respect to measurements performed on the first qubit.}
\label{fig:4s}
\end{figure}

We now give a brief example of how the conversion gate can be used to
generate quantum correlations different from entanglement. Here, the
quantum discord~\cite{Hen,Oll,Mod} allows the quantification of
nonclassical correlations that exist between quantum systems even when
the entanglement is zero. Recently it has been shown that discord can be
used as a resource for various types of quantum protocols that are
useful in a quantum network scenario, including remote state
preparation~\cite{Dakic}, encoding information that only coherent
interactions can extract~\cite{Gu} and verification that untrusted
parties can implement entanglement operations~\cite{Alm}. Discord
represents the quantum component of correlations between two systems,
$A$ and $B$, and is defined as $\delta(A|B)=I(A,B)-J(A|B)$
 with $I(A,B)$ and $J(A|B)$ given by
\begin{eqnarray}
I(A,B)&=& S(\rho_A)+S(\rho_B)-S(\rho_{AB}) \\
J(A|B)&=& S(\rho_A)-\min_{\{\Pi_b\}}\sum p_b S(\rho_{A|b}) \nonumber
\end{eqnarray}
where $S(\rho)$ is the von Neumann entropy $S(\rho)=-{\rm Tr}\rho {\rm
log} \rho$, $\rho_A={\rm Tr}_B \rho_{AB}$, $\rho_B={\rm Tr}_A \rho_{AB}$, $\{\Pi_b\}$ is a positive
operator valued measure on system $B$,
and $p_b = {\rm Tr} [\rho \Pi_b]$ is the probability of obtaining measurement outcome $b$ that
leaves $A$ in the conditional state $\rho_{A|b}$. In general the discord
is not symmetric, $\delta(A|B)\neq \delta(B|A)$, but if it is non-zero with respect to measurements on at least one
system, then nonclassical correlations exist. In addition, when the
entanglement is zero and the discord is non-zero with respect to
measurements on one or both of systems $A$ and $B$ ($\delta(B|A)\neq 0$ or $\delta(A|B)
\neq 0$), then there are quantum correlations present that are not due
to entanglement. 
In order to show that the conversion gate generates such quantum
correlations from initial product states without entanglement we input the state $\rho_{in}=\frac{1}{2}\openone \otimes \ket{+}\bra{+}$. In Fig.~\ref{fig:4s}~(a) the success probability is shown as the angles $\theta_1$ and $\theta_2$ of the gate are modified. In Fig.~\ref{fig:4s}~(b) the corresponding entanglement generated by the gate is shown, as quantified by the concurrence. 

In Figs.~\ref{fig:4s}~(c) and ~\ref{fig:4s}~(d), we give
the discords $\delta(A|B)$ and  $\delta(B|A)$  generated in the output state
with respect to measurements performed on the second and first qubit in
the conversion gate, where $A$ corresponds
to the first qubit in output mode 5 and $B$ corresponds to the second qubit in output mode 6. When $\theta_1=0$ and $\theta_2=\pi/3$ (point (i)) or
$\theta_1=\pi/3$ and $\theta_2=0$ (point (ii)) the output state from the
gate has no entanglement in Fig.~\ref{fig:4s}~(b), but interestingly has non-zero discord (See in Fig.~\ref{fig:4s}~(c)). The output state for these
points is $\frac {5}{7}\ket {H}\bra{H} \otimes
\ket{\phi}\bra{\phi}+\frac {2}{7}\ket {V}\bra{V} \otimes
\ket{+}\bra{+}$, where $\ket{\phi}=(2\ket{H}-\ket{V})/\sqrt 5$ and has a discord of $\delta(A|B)\simeq 0.082$.
The success probability is 0.438 and a Monte Carlo simulation with a $\pm 10$\% variation on the wave plate angles gives $p_s=0.446 \pm 0.023$.

This way the conversion gate can
also be used to generate states with quantum correlations that are not
due to entanglement and may be used for various quantum tasks in a
network scenario~\cite{Dakic,Gu,Alm}, 
most notably as resources in quantum cryptography~\cite{Pirandola}.


\section{State conversion}
One of the powerful features of the nonlocal gate is the ability
to convert quantum states belonging to one type of state into
another that is not LOCC equivalent.

\begin{figure}[tb]
\includegraphics[scale=1]{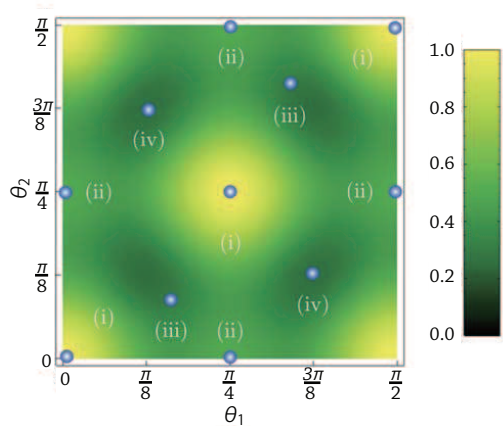}
\caption{The success probability for nonlocal state conversion from the
 linear cluster state $\ket{\rm C_4}$ to various four-qubit entangled
 states. See Tab.~\ref{table:1s} for the labelled states and their success probabilities.}
\label{fig:5s}
\end{figure}

Starting with a four-qubit linear cluster state given by
\begin{eqnarray}
\ket{{\rm C}_4} &=& \frac{1}{2}(\ket{HHHH}+\ket{HHVV}\nonumber\\&&+\ket{VVHH}-\ket{VVVV}),\nonumber
\end{eqnarray}
we will show how the action of the conversion gate shown in Fig.\
\ref{fig:1s} on two qubits of the state $\rho_{in}=\ket{\rm C_4}\bra{\rm
C_4}$ will prepare various LOCC inequivalent states, like a four-qubit GHZ
and Dicke state, as well as into two bipartite maximally entangled states, {\it i.e.} a product of Bell states.   
In general, applying $E_0$ given in Eq.~(\ref{eq:10}) on the second and third qubits of $\ket{\rm C_4}$, we find the general output
\begin{eqnarray}
E_0\ket{\rm C_4}&=&\frac{1}{2} \big[(\alpha_1-\beta_1)\ket {HHHH}-(\alpha_2-\beta_2)\ket {VVVV}\nonumber\\
&&+\mu_1\ket{HHVV}+\mu_1\ket{VVHH}\nonumber\\
&&-\mu_2\ket{HVHV}-\mu_2\ket{VHVH}\big],  \label{eq:22xyy}
\end{eqnarray}
which by tuning the different coefficients enables the generation of a
variety of states as described in the next subsections. In Fig.\
\ref{fig:5s}, we show a selection of output states that can be
obtained by operating the conversion gate on qubits 2 and 3 of
the cluster state together with their success probabilities.


\subsection{Transforming the cluster state into a GHZ state}
In order to obtain a four-qubit GHZ state, $\ket{{\rm GHZ}_4} = (\ket {HHHH}+\ket
{VVVV})/\sqrt 2$, from $\ket{\rm C_4}$ we should design the evolution
operator such that it discards the $\ket{HHVV}$ and $\ket{VVHH}$
components of $\ket{\rm C_4}$ and flips the minus sign in front of the
$\ket{VVVV}$ component. This can be achieved with an operation element
with components $\ket{HH}\bra{HH}$ and $\ket{VV}\bra{VV}$. Thus, for the
output state in Eq.~(\ref{eq:22xyy}) to be $\ket{\rm GHZ_4}$, we should
set $\alpha_j$ and $\beta_j$ such that
$(\alpha_2-\beta_2)=-(\alpha_1-\beta_1)$ and $\mu_j=0$. The latter
equality is satisfied for
$(\theta_1,\theta_2)=(m\frac{\pi}{2},(2n+1)\frac{\pi}{4})$ or
$(\theta_1,\theta_2)=((2n+1)\frac{\pi}{4},m\frac{\pi}{2})$ for
non-negative integers $m$ and $n$. Substituting these in the former
equality leads to $\alpha_1-\beta_1=\pm 1$ and $\alpha_2-\beta_2=\mp
1$. Choosing these coefficients yields 
the Kraus operator of the evolution given by
\begin{eqnarray}
E_0= \pm(\ket {HH}\bra {HH}-\ket {VV}\bra {VV}) \label{eq:22x}
\end{eqnarray}
which achieves the task of transforming  $\ket{\rm C_4}$ to $\ket{\rm GHZ_4}$ (up to a global phase) with a success probability of $p_{s}=1/2$. 
A Monte Carlo simulation with a $\pm 10$\% variation on the wave plate angles gives a success probability of $p_s=0.457 \pm 0.027$.

\subsection{Identity operator: Keeping the cluster state intact} 

If we do not wish to convert $\ket{\rm C_4}$ to any other state but keep
it intact, while still using the nonlocal gate (for practical reasons
re-routing the photons may not be a straightforward procedure), we see
from Eq.~(\ref{eq:22xyy}) that in order to obtain $E_0=\openone$ the
coefficients $\mu_2=0$ and $\alpha_1-\beta_1=\alpha_2-\beta_2=\mu_1 =\pm
1$ are
to be satisfied for $p_s = 1$.
The former equality has
the solutions
$(\theta_1,\theta_2)=(m\frac{\pi}{2},n\frac{\pi}{2})$,
with $m$ and $n$ as non-negative integers. We find $\alpha_j-\beta_j=1$
and $\mu_1=(-1)^{n+m}$ from which we
obtain $\theta_j=k\pi$ and $\theta_j=(2k+1)\frac{\pi}{2}$, respectively,
for even and odd $m$. Then the output state of
Eq.~(\ref{eq:22xyy}) will be equivalent to $\ket{\rm C_4}$ if $n+m$ is
an even number with a success probability of $p_s = 1$. With a $\pm 10$\% variation on the wave plate angles, $p_s=0.909\pm0.056$. 
In this case, the Kraus operator for the successful transformation is the desired identity operator
\begin{eqnarray}
E_0 &=&\ket {HH}\bra {HH}+\ket {VV}\bra {VV}\nonumber\\&& +\ket {HV}\bra {HV}+\ket {VH}\bra {VH}. \nonumber
\end{eqnarray}


\subsection{Transforming the cluster state into two Bell states} 

With a proper choice of the coefficients, we can
disentangle the four qubit entanglement in the cluster state to prepare
two Bell states, one shared between modes 2 and 3, and the other between
the modes 1 and 4. Note that only
the photons in modes 2 and 3 enter the conversion gate. It is easy to
see that among all possible settings of the coefficients the setting
$\alpha_1-\beta_1=\alpha_2-\beta_2=0$ and $\mu_2=-\mu_1$ leads to the desired transformation. Imposing these conditions results in
\begin{eqnarray}
E_0\ket{\rm C_4}&=&\frac{\mu_1}{2}(\ket{HHVV}+\ket{VVHH}\nonumber \\&&+\ket{HVHV}+\ket{VHVH})\nonumber \\
&=&\mu_1\ket{\Psi^+}_{14}\ket{\Psi^+}_{23}.\label{eq:88xyy}
\end{eqnarray}
Thus, if we can find the angles satisfying the above expressions for the coefficients,
 the desired task will be accomplished with the success probability
 $p_s=\mu_1^2$. Here we note that the operation required is the same as the one of the first example given in Section II B. Thus, substituting the angles for this example into the expressions given in Section II B gives $\mu_1=-\mu_2=\pm 1/2$. The success probability is $p_s=\mu_1^2=\frac{1}{4}$. With a $\pm 10$\% variation on the wave plate angles, $p_s=0.270\pm0.017$. The Kraus operator that performs this task is
\begin{eqnarray}
E_0 &=& \frac{1}{2}[\ket {HV}\bra {HV}+\ket {VH}\bra {HV} + \ket { VH}\bra {VH}\nonumber \\ &&+\ket {HV}\bra {VH}]. \nonumber
\end{eqnarray}

\begin{table}[t]
\begin{ruledtabular}
\begin{tabular}{|r|r|r|r|}\hline
{\it Converted state} & $\rm \theta_1$ & $\rm \theta_2$ & \rm $p_s$  \\ \hline
(i) Cluster state& 0 & 0 & 1   \\ 
& $\pi/2$~ &$\pi/2$ & 1 \\ \hline (ii) GHZ state & 0 ($\pi/2$)~ & $\pi/4$ & $1/2$   \\
& $\pi/4$~ & 0 ($\pi/2$) & $1/2$ \\ \hline
(iii) Dicke state & $\rm \theta_{\rm +}$~ &$\rm \theta_{\rm -}$ &$3/10$  \\
 & $\rm \theta_{\rm -}$~ &$\rm \theta_{\rm +}$ & $3/10$ \\ \hline 
(iv) Two Bell states &
 $3\pi/8$~
 &$\pi/8$ & $1/4$  \\
& $\pi/8$~ &$3\pi/8$ & $1/4$  \\ \hline
\end{tabular}
\end{ruledtabular}
\caption{The success probability for nonlocal state conversion from the
 linear cluster state $\ket{\rm C_4}$ to a four-qubit GHZ and Dicke and
 two Bell states. The angle $\theta_{\pm}$ is found from the relation, $\rm \sin 2\theta_{\pm}=\rm \sqrt {\rm (5\pm\sqrt 5)/10}$.\label{table:1s}}
\end{table}

\subsection{Transforming the cluster state into a Dicke state} 

Finally, we show that the nonlocal gate can be used to convert the
linear cluster state $\ket{\rm C_4}$ into the four-qubit Dicke state,
\begin{eqnarray}
\ket{\rm D_4^{(2)}}&& =\frac{1}{\sqrt{6}}(\ket{HVVH}+\ket{VHHV}+\ket{HVHV}\nonumber\\
&&+\ket{VHVH}+\ket{HHVV}+\ket{VVHH}).\label{eq:23www}
\end{eqnarray}
Consider rotating the polarization of the qubits in this state locally
via the operation $\sigma_z\otimes \sigma_x\otimes\sigma_z\sigma_x \otimes
\openone$ to give the following state,
\begin{eqnarray}
\ket{\rm {D'}_4^{(2)}}&& =\frac{1}{\sqrt{6}}(\ket{HHHH}+\ket{VVVV}-\ket{HHVV}\nonumber\\
&&-\ket{VVHH}+\ket{HVHV}+\ket{VHVH}).\label{eq:23www}
\end{eqnarray}
It is now clear that if we can set
$\beta_2-\alpha_2=\alpha_1-\beta_1=-\mu_1=\mu_2$ in
Eq.~(\ref{eq:22xyy}), the coefficients will be equal and the final state
will be locally equivalent to a Dicke state. From $\mu_1=-\mu_2$, we find
$\theta_1-\theta_2=(2k+1)\frac{\pi}{4}$ which leads to the relations
$\sin2\theta_1=(-1)^k\cos2\theta_2$ and
$\cos2\theta_1=(-1)^{k+1}\sin2\theta_2$. Using these relations in
$\alpha_2-\beta_2=\mu_1$, we obtain
$5\sin^42\theta_2-5\sin^22\theta_2+1=0$, whose roots satisfy
$\sin^22\theta_2=(5\pm\sqrt{5})/10$. Using the expression for
$\sin^22\theta_2$ in $\sin^22\theta_1+\sin^22\theta_2=1$, which is
derived from the equality $\alpha_2-\beta_2=\beta_1-\alpha_1$, we find
$\sin^22\theta_1=(5\mp\sqrt{5})/10$. Then, setting
$\sin^22\theta_1=(5\pm\sqrt{5})/10$ gives
$\sin^22\theta_2=(5\mp\sqrt{5})/10$ and leads to
$\alpha_1-\beta_1=\mp1/\sqrt{5}$ and
$\alpha_2-\beta_2=\mu_1=\mu_2=\pm1/\sqrt{5}$. Inserting these values for the
coefficients in Eq.~(\ref{eq:22xyy}), we arrive at
\begin{eqnarray}
E_0\ket{\rm C_4}&&=\mp\frac{1}{2\sqrt{5}}[\ket {HHHH}+\ket {VVVV} -\ket{HHVV}\nonumber\\
&&\qquad \qquad-\ket{VVHH}+\ket{HVHV}+\ket{VHVH}]\nonumber\\&&=\mp \frac{\sqrt{6}}{2\sqrt{5}}\ket{\rm {D'}_4^{(2)}} \label{eq:22xyzz}
\end{eqnarray}
implying that we can convert $\ket{\rm C_4}$ into $\ket{\rm {D}_4^{(2)}}$ with a success probability $p_s=3/10$. 
The Monte Carlo simulation with a $\pm 10$\% variation on the wave plate angles gives a success probability of $p_s=0.303 \pm 0.013$.
The Kraus operator for the successful transformation is given by
\begin{eqnarray}
E_0&=&\mp\frac{1}{\sqrt{5}}(\ket {HH}\bra {HH}-\ket {VV}\bra {VV}-\ket {HV}\bra {HV} \nonumber \\
&&-\ket{VH}\bra {VH} +\ket {VH}\bra{HV}+\ket {HV}\bra {VH}).\nonumber
\end{eqnarray}
\vskip0.5cm


\begin{figure*}[t]
\includegraphics[width=17cm]{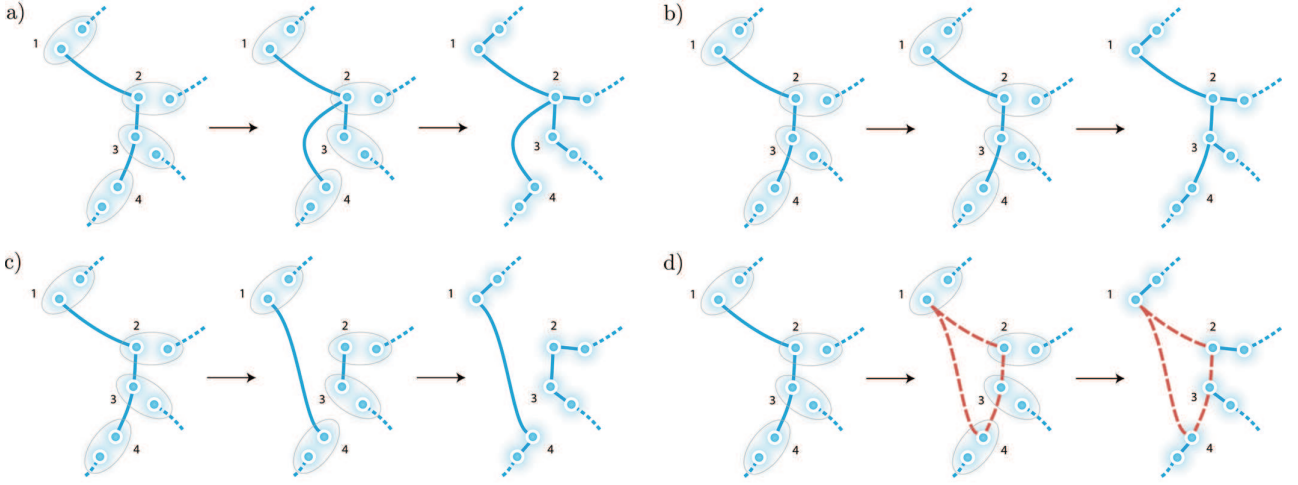}
\caption{Nonlocal gate used in a multi-qubit network taking the form of a graph state. Here, vertices are qubits initialized to the state $\frac{1}{\sqrt{2}}(\ket{H}+\ket{V})$ and edges correspond to the application of a controlled-$Z$ ($CZ$) operation between the vertices: $CZ=\ket{H}\bra{H} \otimes \openone +\ket{V}\bra{V} \otimes \sigma_z$. In the first step of each panel, a linear cluster state has local operations ${\sf H} \otimes \openone \otimes \openone \otimes {\sf H} $ applied (${\sf H}$ is the Hadamard operation) to take it from its canonical graph state form to the state $\ket{{\rm C}_4}$. It is then converted into various four-qubit entangled states using the nonlocal gate, as summarized in Tab.~\ref{table:1s}. In the second step of each panel, these states are then connected to the rest of the graph representing the network via $CZ$ gates (and local operations). See main text for details. The dashed edges represent $CZ$ gates applied between the qubits shown and other qubits in the total graph (not shown). {\bf (a):}~Cluster state converted into a star graph (locally equivalent to the GHZ state), which is then connected to the rest of the graph of the network. {\bf (b):} Cluster state remains intact as a cluster state and is connected to the graph (after local operations). {\bf (c):} Cluster state converted into two 2-qubit graph states (each locally equivalent to a Bell state). {\bf (d):} Cluster state converted into a Dicke state and connected to the graph to form a hybrid quantum network. Here, the red dashed edges signify the state is entangled and are not $CZ$ operations.}
\label{multi}
\end{figure*}

\section{Nonlocal gate in a multi-qubit network}

We now briefly discuss the integration of the nonlocal gate in a multi-qubit network in the form of a graph state. Graph states can be used for a wide variety of quantum networking purposes, including in quantum communication and distributed quantum computation~\cite{clusterrev,Hein}. In this setting an important task is to `rewire' the network by changing the entanglement structure in order to carry out a specific protocol between selected parties. From the previous section it is clear that the nonlocal gate can convert a cluster state into a number of entangled states with different entanglement structures, as summarized in Tab.~\ref{table:1s}. We now show that as a result the nonlocal gate enables the rewiring of a graph state network. 

Consider the initial cluster state as part of a multi-qubit graph state, as depicted in the first step of Fig.~\ref{multi}~(a). Here, the cluster state is shown in its canonical graph state form, where vertices correspond to qubits in the state $\frac{1}{\sqrt{2}}(\ket{H}+\ket{V})$ and solid edges correspond to the application of a controlled-$Z$ ($CZ$) operation between the vertices: $CZ=\ket{H}\bra{H} \otimes \openone +\ket{V}\bra{V} \otimes \sigma_z$. The dashed edges represent $CZ$ gates applied between the qubits shown and other qubits in the total graph state (not shown). The canonical cluster state can be converted into the state $\ket{{\rm C}_4}$ with local operations ${\sf H} \otimes \openone \otimes \openone \otimes {\sf H} $, where ${\sf H}$ is the Hadamard operation. Once these local operations have been performed, the nonlocal gate is then applied. Here, there are four rewiring cases:

{\it (a) Rewiring into a star cluster.--} The nonlocal gate is applied between qubits 2 and 3 for the cluster state $\ket{{\rm C}_4}$ to become a GHZ state. This state is equivalent to the star cluster state shown in step 2 of Fig.~\ref{multi}~(a) under local operations ${\sf H} \otimes \openone \otimes {\sf H} \otimes {\sf H} $. Once these bilocal operations have been performed, $CZ$ operations are then applied to connect the star cluster state into the total graph state, as shown in step 3. This rewires the network. Note that one can choose any qubit to be the central node of the star cluster in step 2 ($\openone$ is applied to the central node and ${\sf H}$ to the outer nodes), giving four possible rewiring configurations. It is also interesting to note that the star cluster state and linear cluster state are not equivalent under local operations and classical communication~\cite{Hein}, making the use of the nonlocal conversion gate necessary in order to rewire the network in this case.

{\it (b) Keeping the initial wiring.--} The nonlocal gate is applied between qubits 2 and 3 of the cluster state $\ket{{\rm C}_4}$, which remains as $\ket{{\rm C}_4}$. This state is equivalent to the linear cluster state shown in step 2 of Fig.~\ref{multi}~(b) under local operations ${\sf H} \otimes \openone \otimes \openone \otimes {\sf H} $. Once these local operations have been performed, $CZ$ operations are then applied to connect the linear cluster state into the total graph state. This keeps the initial wiring of the network. 

{\it (c) Rewiring into two graphs.--} The nonlocal gate is applied between qubits 2 and 3 of the cluster state $\ket{{\rm C}_4}$, which becomes a product of two Bell states. Each of these states are equivalent to a two-qubit graph state, as shown in step 2 of Fig.~\ref{multi}~(c), under local operations $\openone \otimes {\sf H}\sigma_x$. Once these local operations have been performed, $CZ$ operations are then applied to connect the graph states into the total graph state. This rewires the network.

{\it (d) Rewiring into a hybrid network.--} The nonlocal gate is applied between qubits 2 and 3, and the corresponding local operations are performed, for the cluster state $\ket{{\rm C}_4}$ to become a Dicke state $\ket{{\rm D}^{(2)}_4}$. This state is shown in step 2 of Fig.~\ref{multi}~(d), where the internal entanglement connections of the Dicke state (not $CZ$ gates) are represented by dashed red edges. $CZ$ operations are then applied to connect each qubit of the Dicke state individually into the total graph state. This is a standard method for encoding qubits into graph states~\cite{clusterrev}. The operations produce a hybrid network of a graph state and a Dicke state that may be more efficient for certain distributed protocols, such as in telecloning and quantum secret sharing~\cite{Prevedel}.


\section{Conclusion}
In this work we proposed a simple and powerful linear optical quantum gate that can be used
to prepare entanglement, or to observe states without entanglement but nonzero discord. In addition, we showed how to convert a four-qubit linear
cluster state into a series of relevant multipartite entangled states, such as a
four-qubit GHZ, Dicke state and two bipartite maximally entangled states,
by acting on only two qubits. The gate
can perform a range of nonlocal operations based on its functionality as
a tunable polarization dependent beamsplitter. As a result, it can be used to implement various different types of fusion operation in a number of quantum state expansion schemes \cite{cluster1,cluster2,s18}. Indeed, recent work has also shown how to scale quantum
networks by fusing small multipartite entangled states containing
four-qubit entanglement \cite{Fujii}.Thus, by placing the gate in either
a large-scale quantum network \cite{Kimble} or small-scale on-chip network \cite{Silverstone}, we envisage that it could play an important role
in efficiently preparing and converting other types of larger
multipartite entangled states where there may be restricted access to a
given resource.

\acknowledgements
This work was supported by a JSPS Postdoctoral Fellowship for Research
Abroad. This work was also supported by EU-FET project QWAD, the German excellence research cluster NIM, the South African National Research Foundation and the South African National Institute for Theoretical Physics. TT is supported by Grant-in-Aid for Young Scientists (B), no.15K17729. FN was partially supported by the RIKEN iTHES Project, the MURI Center for Dynamic Magneto-Optics, the IMPACT program of JST, a Grant-in-Aid for Scientific Research (A), and a grant from the John Templeton Foundation.

\end{document}